\documentclass[reprint,amsmath,amssymb,pra,twocolumn,longbibliography]{revtex4-1}

\usepackage{graphicx,multirow,color}
\usepackage{csquotes}
\usepackage{bm}

\newcommand\opex{Opt. Express}

\begin{document}
\title{Perturbation theories for symmetry-protected bound states in the continuum on two-dimensional periodic structures}

\author{Lijun Yuan}
\email{Corresponding author: ljyuan@ctbu.edu.cn}
\affiliation{College of Mathematics and Statistics, Chongqing Technology and Business University, Chongqing, 
China}
\author{Ya Yan Lu}
\affiliation{Department of Mathematics, City University of Hong Kong, Hong Kong}

\begin{abstract}
On dielectric periodic structures with a reflection symmetry in a
periodic direction, there can be antisymmetric standing waves (ASWs)
that are symmetry-protected bound states in the continuum (BICs). The
BICs have found many applications, mainly because they give rise to
resonant modes of extremely large quality-factors ($Q$-factors). 
The ASWs are robust to symmetric  perturbations of the structure, but
they become resonant modes if the perturbation is non-symmetric. The
$Q$-factor of a resonant mode on a perturbed
structure is typically $O(1/\delta^2)$ where $\delta$ is the amplitude
of the  perturbation, but special perturbations can produce resonant
modes with larger $Q$-factors.  
For two-dimensional (2D) periodic structures with a 1D periodicity, 
we derive conditions on the perturbation profile such that the $Q$-factors are $O(1/\delta^4)$ or
$O(1/\delta^6)$. For the unperturbed structure, an ASW is surrounded
by resonant modes with a nonzero Bloch wave vector. For 2D periodic
structures, the $Q$-factors of nearby resonant modes are typically
$O(1/\beta^2)$, where $\beta$ is the Bloch wavenumber. We show that
the $Q$-factors can be $O(1/\beta^6)$ if the ASW satisfies a simple
condition.  
\end{abstract}

\maketitle

\section{Introduction}

In recent years, bound states in the continuum (BICs) have attracted
much attention in the photonics community \cite{neumann29,hsu16,kosh19}.
In classical wave systems, a BIC is a trapped or guided mode with a
frequency in the frequency interval where outgoing 
radiative waves exist. Mathematically, a BIC corresponds to a discrete
eigenvalue in the continuous spectrum, and causes the non-uniqueness
of a corresponding scattering or diffraction problem
\cite{bonnet94,shipman03}.  Optical BICs have been studied in a number
configurations including waveguides with local distortions
\cite{evans94,bulg08}, waveguides with
lateral leaky structures
\cite{plot11,weim13,zou15}, waveguides with
anisotropic materials \cite{gomis17}, and periodic structures surrounded by or
sandwiched between homogeneous media
\cite{bonnet94,shipman03,padd00,tikh02,shipman07,lee12,hu15,port05,mari08,hsu13_1,hsu13_2,yang14,bulg14b,bulg15,gan16,li16,gao16,ni16,yuan17,bulg17,hu17josab,han18,hanD19}.
On symmetric structures, there may be symmetry-protected BICs that
do not couple with the radiative waves due to a symmetry mismatch
\cite{bonnet94,shipman03,padd00,tikh02,shipman07,lee12,hu15}.  Other
BICs exist without any symmetry mismatch, but in some cases, their
existence still depends crucially on the symmetry and they are robust
against structural perturbations that preserve the symmetry
\cite{zhen14,bulg17pra,yuan17_4}.  Optical BICs have applications in
lasing \cite{kodi17}, sensing \cite{romano19,yesi19}, filtering
\cite{foley14,cui16}, and switching \cite{han19}, and for
enhancing emissive processes and nonlinear optical effects
\cite{yuan16,yuan17_2}. 

A BIC can be considered as a resonant mode with an infinite
$Q$-factor. This implies that resonant modes with extremely large
$Q$-factors can be created by perturbing the structure or varying a
physical parameter slightly \cite{yuan17_2,bulg17_2,yuan18,
  kosh18,hu18}. Since many applications of the BICs are related to the
high-$Q$ resonances, it is of significant importance to understand how
the $Q$-factors depend on the structural or parametric perturbations.
It is well-known that symmetry-protected BICs are robust with respect
to symmetric perturbations \cite{yuan17_4}. For a non-symmetric
perturbation of amplitude $\delta$, the symmetry-protected BICs become
resonant modes and their $Q$-factors are typically $O(1/\delta^2)$
\cite{kosh18}. On periodic structures, a BIC is always surrounded by a
family of resonant modes depending on the wave vector. For 2D periodic
structures with a 1D periodicity, if there is a BIC with a Bloch
wavenumber $\beta_*$, then the resonant mode with Bloch wavenumber
$\beta$ (close to $\beta_*$) typically has a $Q$-factor proportional to
$1/|\beta - \beta_*|^2$.  In previous works \cite{yuan17_2,yuan18}, we
identified some special BICs such that
the $Q$-factors of the nearby resonant modes are at least $O(1/|\beta -
\beta_*|^{4})$.  It is also known that for some cases, the $Q$-factors
are $O(1/|\beta-\beta_*|^6)$ \cite{jin19}.

In this paper, we present new perturbation results for antisymmetric
standing waves (ASWs) on 2D periodic structures with a reflection
symmetry in the periodic direction. The ASWs are symmetry-protected BICs
since they have a symmetry mismatch with plane waves propagating in
the normal direction (perpendicular to the periodic direction). 
For a non-symmetric structural perturbation with amplitude $\delta$ and 
profile $F$, we show that the $Q$-factor is in general $O(1/\delta^2)$, but
 it becomes $O(1/\delta^4)$ if $F$ satisfies a simple
condition involving the wave field of the ASW, and it further becomes
$O(1/\delta^6)$ if in addition to the previous 
condition, $F$ is anti-symmetric. We also consider the 
perturbation with respect to the Bloch wavenumber. 
It is known that for any BIC (with Bloch wavenumber $\beta_*$) 
satisfying a special condition \cite{yuan18}, the $Q$-factors of nearby 
resonant modes (with Bloch wavenumber $\beta$)  are $O( 1/|\beta -
\beta_*|^{4})$ in general. We show that if the BIC is an ASW, then 
the $Q$-factors are in fact $ O(1/\beta^{6})$.
These theoretical results are validated by numerical
examples for periodic structures with three arrays of circular
dielectric cylinders. 

The rest of this paper is organized as follows. In
Sec.~\ref{sec:SPBIC},  a brief introduction is given for
BICs and resonant modes on 2D periodic structures. In
Sec.~\ref{sec:material_pert}, a detailed structural perturbation
theory is developed for ASWs on periodic structures with a reflection
symmetry. In Sec.~\ref{sec:beta_pert}, a special wavenumber perturbation
theory  is briefly presented for ASWs. 
To validate the perturbation theories,   numerical examples are  
presented in Sec.~\ref{sec:example}.  The paper is concluded with a
brief discussion in Sec.~\ref{conclusion}.

\section{BICs and resonant modes}
\label{sec:SPBIC}

In this section, we present the mathematical formulation and give
precise definitions for BICs and resonant modes. 
We consider 2D dielectric structures that are invariant in $z$, periodic in $y$ with
period $L$, and bounded in the $x$ direction by $|x|<D$ for some
constant $D$, where $\{x,y,z\}$ is a Cartesian coordinate system. 
Assuming the surrounding medium for $|x| > D$ is vacuum, we have a 
dielectric function $\epsilon$ satisfying
$\epsilon(x,y+L)=\epsilon(x,y)$ for all $(x,y)$, and
$\epsilon(x,y)=1$ for $|x| > D$. For the $E$ polarization, 
the $z$-component of the electric field, denoted as $u$,
satisfies the following Helmholtz  equation 
\begin{equation}
\label{eq:helm} \frac{\partial^2 u}{\partial x^2} + \frac{\partial^2
  u}{\partial y^2} + k^2 \epsilon u = 0, 
\end{equation}
where $k = \omega/c$ is the free space wavenumber, $\omega$ is the
angular frequency, $c$ is the speed of light in vacuum, and the
time dependence is assumed to be $e^{- i \omega t}$. 

A Bloch mode on such a periodic structure is a solution of Eq.~(\ref{eq:helm}) given as
\begin{equation}
\label{eq:bloch} u(x,y) = \phi(x,y) e^{i \beta y}, 
\end{equation}
where $\phi$ is periodic in $y$ with period $L$ and $\beta$ is the
real 
Bloch wavenumber. Due to the periodicity of $\phi$, $\beta$ can be
restricted to the interval $[-\pi/L, \pi/  L]$. 
If $\omega$ is real and $\phi \to 0$ as $|x| \to \infty$, then $u$ given by
Eq.~(\ref{eq:bloch}) is a guided mode. 
Typically, guided modes that depend 
on $\beta$ and $\omega$ continuously  can only be found below the
light line, i.e., for $ k < |\beta|$. A BIC is a special guided mode above the light line, i.e., $\beta$
and $k$ satisfy the condition $k> |\beta|$. 
For a given structure, BICs can only exist at isolated points in the
$\beta$-$\omega$ plane. 

In the homogeneous medium given by $|x| > D$, we can expand a Bloch mode in
plane waves, that is 
\begin{equation}
\label{planewave} u(x,y) = \sum_{j = - \infty}
^{\infty}c_{ j}^{\pm} e^{  i (\beta_j y \pm  \alpha_ j x)}, \quad |x|>
D, 
\end{equation}
where the ``$+$'' and ``$-$'' signs are chosen 
for $x > D$ and $x < -D$ respectively, and 
\begin{equation}
  \label{defab}
\beta_j = \beta + \frac{2\pi j}{L}, \quad 
\alpha_j = \sqrt{k^2 - \beta_j^2}.  
\end{equation}
If $\omega$ is positive and $k < |\beta_j|$, then $\alpha_j = i \sqrt{\beta_j^2 - k^2}$ is pure
imaginary, and the corresponding plane wave is evanescent.
 For a BIC, one or more $\alpha_j$ are real,  and the
corresponding coefficients $c_j^\pm$ vanish, since the field must
decay to zero as $|x| \to \infty$. 

On periodic structures with a reflection symmetry in the $y$
direction, there may be ASWs which are  symmetry-protected BICs
with $\beta = 0$. Assuming the dielectric function $\epsilon(x,y)$ is
even in $y$,  then the field $u$ of an ASW is an odd function of
$y$. If the ASW is non-degenerate, i.e., the
  multiplicity is 1, the real and imaginary parts of $u$ (if they
  are nonzero) are 
themselves ASWs of the same frequency, thus they are 
linearly dependent. Without loss of generality,  we assume the field
$u$ of a non-degenerate ASW is real.
We consider only ASWs with a frequency $\omega$  such that $0 < k
< 2 \pi / L$. In that case, $\alpha_0 = k$ is real and all other $\alpha_j$ for $j
\neq 0$ are pure imaginary. Since $u$ is odd in $y$, 
the coefficients $c_0^{\pm}$ are always zero.

On periodic structures, a resonant mode is a Bloch mode with a complex
frequency $\omega$ and a real Bloch wavenumber $\beta$ and it radiates
out power as $|x| \to \infty$.  
The resonant modes form bands, and on each band, they depend on
$\beta$ continuously. Under the assumed time dependence, the imaginary
part of the complex 
frequency of a resonant mode must be negative, so that its amplitude
can decay with time. The $Q$-factor of a resonant mode can be defined
as $Q = -0.5 
\mbox{Re}{(\omega)}/\mbox{Im}(\omega)$. The expansion
(\ref{planewave}) 
is still valid, but the complex square root for $\alpha_j$ must be defined
to maintain continuity as $\mbox{Im}(\omega) \to 0$. This can be 
achieved by using a square root with a branch cut along the negative
imaginary axis, that is, if $\xi = |\xi| e^{ i \theta}$ for $-\pi/2 <
\theta \le 3\pi/2$, then $\sqrt{\xi} = \sqrt{|\xi|} e^{i \theta/2}$. 
For a resonant mode above the light line, i.e. $\mbox{Re}(k)
> |\beta|$,   $\alpha_0$ has a negative imaginary part, and the field 
blows up as $|x| \to \infty$.

\section{Structural perturbation} 
\label{sec:material_pert}


In this section, we analyze the effects of structural perturbations to
ASWs on 2D periodic structures with a reflection symmetry in the
periodic direction. 
 Let $u_*$ be a non-degenerate ASW with a positive frequency
$\omega_*$ on a periodic  
structure given by a dielectric function $\epsilon_*$, where 
$\epsilon_*$ is even in $y$ and $u_*$ is odd in $y$. We assume 
$k_* =\omega_*/c$ satisfies
\begin{equation}
  \label{cutoff1}
0 < k_* <   2\pi/ L, 
\end{equation}
and consider a perturbed structure  given by 
\begin{equation} 
\label{eq:pert_eps} \epsilon(x,y) = \epsilon_*(x,y) + \delta F(x,y),
\end{equation}
where $\delta$ is small real parameter, and $F$ is the real 
perturbation profile satisfying $F(x,y) = 0$ for $|x| > D$ and  $\max
|F| = 1$. If $F$ is even in $y$,  the 
perturbation can only shift the frequency on the real line, thus
the perturbed structure still has an
ASW near the original one \cite{yuan17_4}. In the following, we assume $F$
is non-symmetric, i.e., not an even function of $y$. 

We are interested in the complex frequency
$\omega$ and the $Q$-factor of a nearby resonant mode $u$ with Bloch
wavenumber $\beta= 0$ on the perturbed structure. 
In the perturbation analysis, we expand $u$ and
 $ \omega $ as
\begin{eqnarray}
\label{u_exp} u &=& u_{\ast}  + u_1 \delta + u_2 \delta^2 + 
         u_3 \delta^3 + u_4 \delta^4 + \ldots  \\
\label{omg_exp} \omega &=& \omega_* + \omega_1 \delta + \omega_2 \delta^2 + \omega_3
           \delta^3 + \omega_4 \delta^4 + \ldots
\end{eqnarray}
Inserting Eqs.~(\ref{u_exp})-(\ref{omg_exp}) into 
Eq.~(\ref{eq:helm}),  and comparing terms of equal power in $\delta$,
we obtain  
\begin{eqnarray}
  \label{eq:ustar}
  {\cal L} u_* &=& 0, \\
\label{eq:uone}
 {\cal L} u_1 &=&
 - (2 k_* k_1 \epsilon_* + k_*^2 F) u_*,\\
\nonumber
  {\cal L} u_2 &=&  - (2 k_* k_1 \epsilon_* + k_*^2 F) u_1   \\ 
\label{eq:utwo}
&& - (2 k_* k_2 \epsilon_* + k_1^2 \epsilon_* + 2 k_* k_1 F) u_*, 
\end{eqnarray}
where  $k_j = \omega_j / c$ for $j \geq 1$,  and
\begin{equation}
  \label{defL}
{\cal L} = \partial_x^2 + \partial_y^2 
+ k_*^2  \epsilon_*.  
\end{equation}
In addition, $u_j$ must satisfy a proper outgoing radiation 
condition as $|x| \to \infty$ and a periodic condition in the $y$ direction.  

Equation (\ref{eq:ustar}) is simply the governing Helmholtz
equation of the ASW. The first order term $u_1$ satisfies the
inhomogeneous Eq.~(\ref{eq:uone}) which is singular and has no
solution, unless the right hand side is orthogonal to
$u_*$. Let $\Omega$ be the infinite strip given by $|y| < L/2$ and
$-\infty < x < \infty$. 
Multiplying $\overline{u}_*$ (the complex conjugate of
$u_*$) to both sides of Eq.~(\ref{eq:uone}) and integrating on
domain $\Omega$, we obtain 
\begin{equation}
  \label{eq:k1}
k_1  = \frac{\omega_1}{c}  = - \frac{  k_* \int_\Omega F |u_*|^2 \, d{\bm r} }
{2 \int_\Omega \epsilon |u_*|^2 \, d{\bm r} }.
\end{equation}
Since $F$ is real,  it is clear that $k_1$ is also real. 

For $k_1$ given above, Eq.~(\ref{eq:uone}) has a solution. 
 Similar to the plane wave expansion (\ref{planewave}), 
$u_1$ can be written down explicitly for $|x| > D$. 
Importantly, condition (\ref{cutoff1}) implies that $u_1$ contains only a single
outgoing plane wave as $x \to \pm \infty$, that is 
\begin{equation}
\label{u1_exp}
u_1 \sim  b_0^{\pm} e^{\pm i k_{\ast  } x}, \quad x \to \pm
\infty, 
\end{equation}
where $b_0^{\pm}$ are unknown coefficients. A formula for $k_2$ can be derived from the solvability condition of
Eq.~(\ref{eq:utwo}). In particular, the imaginary part of $k_2$ has
the following simple formula
\begin{equation}
  \label{eq:k2_imag}
  \mbox{Im}(k_2) 
  = \frac{\mbox{Im}(\omega_2)}{c} 
= - \frac{L  \left( |b_{0}^+|^2 +|b_{0}^-|^2 \right)} {2 
   \int_\Omega \epsilon |u_*|^2 \, d{\bm r}}.
\end{equation}
 A derivation of Eq.~(\ref{eq:k2_imag}) is given in 
Appendix A. 

Notice that if $u_1$ radiates power to $x=\pm \infty$, $b_0^+$ and
$b_0^-$ are nonzero, then $\mbox{Im}(\omega_2) \ne 0$. In that case, the
imaginary part of the complex frequency satisfies 
\begin{equation}
  \label{imgomg2}
  \mbox{Im}(\omega) \approx   \mbox{Im}(\omega_2) \delta^2 
= 
 - \frac{ c L  \left( |b_{0}^+|^2 +|b_{0}^-|^2 \right)} {2 
    \int_\Omega \epsilon |u_*|^2 \, d{\bm r}} \delta^2, 
\end{equation}
and the $Q$-factor satisfies 
\begin{equation}
  \label{Qorder2}
  Q \approx   \frac {k_{\ast}  \int_\Omega \epsilon |u_*|^2 \, d{\bm r}}
  { L   \left( |b_{0}^+|^2 +|b_{0}^-|^2 \right)}   \delta^{-2}.
\end{equation}


If we can find a perturbation profile $F$ such that $b_0^\pm = 0$, then $\mbox{Im}(\omega_2)=0$, and Eqs.~(\ref{imgomg2})
and 
(\ref{Qorder2}) are no longer valid. In that case,
$\mbox{Im}(\omega_3)$ must also be 
zero,  since otherwise, $\mbox{Im}(\omega)$ changes signs when $\delta$
change signs. This is not possible, since
$\mbox{Im}(\omega)$ of a resonant mode is always negative. Therefore,
if $u_1$ is non-radiative, we expect $\mbox{Im}(\omega) \sim
\delta^4$ and $Q \sim \delta^{-4}$.   In the following, we derive a
condition for $F$ that guarantees $b_0^\pm =
0$.  This condition involves the ASW $u_*$ and 
related diffraction solutions for normal incident waves at the same
$\omega_*$. 

We consider two diffraction problems with normal incident waves 
$e^{ i  k_* x}$ and $e^{- i   k_* x}$
given in the left and right homogeneous media, respectively. The
solutions of these two diffraction problems are denoted as $v_1$ and
$v_2$, respectively. Note that $v_1$ and $v_2$ can be chosen as 
even functions of $y$. 
The existence of a BIC 
implies that the corresponding diffraction problems have no uniqueness 
\cite{bonnet94,shipman03}, but the diffraction solutions are
uniquely defined in the far field as $|x| \to \infty$.
In fact, $v_1$ and $v_2$ have the following
asymptotic formulae
\begin{eqnarray}
\label{leftasym} 
v_1  &\sim& 
\begin{cases}
    e^{ i k_\ast x} + S_{11} e^{-i 
    k_\ast x},   & x\to -\infty \\
  S_{21} e^{i k_{\ast} x },  & x \to +\infty, 
\end{cases} \\
v_2  &\sim & 
\begin{cases}
 S_{12}  e^{ -i k_\ast x},  & x\to -\infty \\
e^{-i k_* x} + S_{22} e^{i k_* x},  & x \to +\infty,
\end{cases}
\end{eqnarray}
where $S_{11}$, $S_{21}$, $S_{22}$ and $S_{12}$ are the reflection and
transmission coefficients associated with the left and right incident
waves, respectively, and they are the entries of a  $ 2\times 2$
scattering matrix  ${\bm S}$. Due to the reciprocity and energy
conservation, ${\bm S}$ is a symmetric and unitary matrix \cite{yuan19}.  

For Eq.~(\ref{eq:uone}), we would like to multiply both sides by
$\overline{v}_m$ (for $m=1$, 2) and integrate on domain $\Omega$. 
 Since $u_* $ decays to zero exponentially as $x \to \pm \infty$, the 
right hand side, i.e., 
\begin{equation*}
G_m = - \int_\Omega (2 k_* k_1 \epsilon_* + k_*^2 F) u_* \overline{v}_m  d{\bm r}
\end{equation*}
is well defined. Since $\epsilon_*$ and $v_m$ are even in $y$, and
$u_*$ is odd in $y$, $G_m$ can be simplified as  
\begin{equation}
  \label{defGG}
G_m = - k_*^2 \int_\Omega  F u_* \overline{v}_m  d{\bm r}. 
\end{equation}
On the other hand, $v_m$ and $u_1$ (in general) do not
decay to zero  as $|x|
\to \infty$, it is not immediately clear whether the right hand side
can be integrated on the unbounded domain $\Omega$. However, for any
$h \ge D$, we can define a rectangular domain 
$\Omega_h$ given by $|y| < L/2$ and $|x| < h$, and evaluate the
integral on $\Omega_h$, then take the limit as $h \to \infty$.
Clearly, the limit must exist and 
\[
\lim_{h \to \infty} \int_{\Omega_h} \overline{v}_m {\cal L} u_1 
d{\bm r} 
=  G_m, \quad m =1, 2. 
\]
In Appendix B, we show that 
\begin{equation}
\label{leftinc}
\lim_{h \to \infty} \int_{\Omega_h} \overline{v}_m {\cal L} u_1 
d{\bm r} = 
2 i k_* L \left( b_0^-   \overline{S}_{1m} + b_0^+ \overline{S}_{2m}
\right). 
\end{equation}
Therefore,
\begin{eqnarray}
\label{coeffu1}
   G_m =  2 ik_* L \left( b_0^-   \overline{S}_{1m} + b_0^+
  \overline{S}_{2m}   \right), \quad m=1, 2.
\end{eqnarray}
Since the scattering matrix ${\bm S}$ is invertible, 
it is clear that $b_0^+ = b_0^- = 0$ if and only if $G_1 = G_2 =
0$. Therefore, if we choose a  perturbation profile $F$ such that
$G_1=G_2=0$, then the  $Q$-factors of resonant modes near the ASW
should satisfy at least an inverse fourth order
asymptotic relation, i.e.  $Q \sim \delta^{-4}$.  
Notice that the condition $G_1=G_2 = 0$ is unchanged,
  if we replace $v_1$ and $v_2$ in Eq.~(\ref{defGG}) by their 
linear combinations $w_1$ and $w_2$, as far as $w_1$ and $w_2$ are 
linearly independent. In Appendix D, we show that $w_1$ and $w_2$ can
be chosen as real functions. Therefore, $G_1=G_2=0$ gives only two
independent real conditions.


Many different perturbation profiles can satisfy the condition
$G_1=G_2 = 0$. 
In the following,  we show that if $G_1=G_2=0$ and $F$  is an odd
function of $y$, then 
the $Q$-factors satisfy an inverse sixth order asymptotic relation, 
i.e.  $Q \sim \delta^{-6}$. 
From the assumptions, it is clear that $k_1=0$, 
$b_0^{\pm} = 0$, $u_1 \to 0$ as $| x| \to \infty$, and $k_2$ is
real. Since the right hand side of
Eq.~(\ref{eq:uone}) is an even function of $y$, we can choose
$u_1$   as  an even function of $y$. The right hand side of
Eq.~(\ref{eq:utwo}) is an odd function of $y$, so $u_2$ can be chosen
as an odd function of $y$.  
 Similar to Eq.~(\ref{u1_exp}), $u_2$ has the asymptotic formulae
\begin{equation}
\label{u2_exp} u_2 \sim b_1^{\pm} e^{\pm k_* x},  \quad x \to \pm \infty,
\end{equation}
where $b_1^{\pm}$ are unknown coefficients. The solvability of 
Eq.~(\ref{eq:uthree}) for $u_3$ leads to  $k_3 = 0$ as shown in
Appendix C.  In principle, $k_4$ can be obtained from the
solvability of Eq.~(\ref{eq:ufour}) for $u_4$. In Appendix C, we show
that 
\begin{equation}
  \label{eq:k4_imag}
  \mbox{Im}(k_4) 
  = \frac{\mbox{Im}(\omega_4)}{c} 
= - \frac{L  \left( |b_{1}^+|^2 +|b_{1}^-|^2 \right)} {2 
   \int_\Omega \epsilon |u_*|^2 \, d{\bm r}}.
\end{equation}


Since $k_1 = 0$, the right hand side of Eq.~(\ref{eq:utwo}) for $u_2$
is simplified to $-  k_*^2 F u_1   - 2 k_* k_2 \epsilon_*  u_*$. 
It decays to zero exponentially as $|x| \to \infty$, and  is an odd
function of $y$. Since the diffraction solutions $v_m$ (for $m=1$, 2)
are even in $y$, we have
\begin{equation}
\label{eq:Hm}
 H_m = - \int_\Omega  (k_*^2 F u_1    + 2 k_* k_2 \epsilon_*  u_* )
 \overline{v}_m d {\bm r} = 0. 
 \end{equation}
Similar to Eq.~(\ref{coeffu1}), we  multiply the left hand side of
Eq.~(\ref{eq:utwo}) by $\overline{v}_m$, integrate on domain
$\Omega_h$, take the limit as $h \to \infty$, and obtain
\begin{equation}
\label{eq:Hm2}
H_m = 2 i k_* L \left( b_1^-   \overline{S}_{1m} + b_1^+ \overline{S}_{2m}   \right), \quad m = 1, 2.
\end{equation}
From Eqs.~(\ref{eq:Hm}) and (\ref{eq:Hm2}) above, we obtain 
$b_1^{-} = b_1^+ = 0$. Therefore,  $\mbox{Im}(\omega_4) = 0$ and $u_2$
does not radiate out power  as $x \to \pm \infty$. 
Meanwhile,  $\mbox{Im}(\omega_5)$ must be zero, since otherwise 
$\mbox{Im}(\omega)$ will have a wrong sign as $\delta$ changes 
signs. Since $k_3 = 0$, the right hand side of Eq.~(\ref{eq:uthree}) for
$u_3$ is even in $y$. 
It is possible for $u_3$ to radiate out power as $x \to
\pm \infty$, then $\mbox{Im}(\omega_6)$ is  nonzero, 
$\mbox{Im}(\omega) \sim \delta^6$, and $Q \sim \delta^{-6}$. 

In summary, for a general real perturbation profile $F$, $Q$ is
typically $O(1/\delta^2)$ and is given in Eq.~(\ref{Qorder2}); if $F$
satisfies the condition $G_1=G_2=0$ for $G_m$ given in
Eq.~(\ref{defGG}), then $Q$ is at least $O(1/\delta^4)$; and if $F$ is
odd in $y$ and $G_1=G_2 = 0$, then $Q$ is at least $O(1/\delta^6)$. 

\section{Wavenumber perturbation}
\label{sec:beta_pert}

A BIC with a Bloch wavenumber $\beta_*$ on a periodic structure is
surrounded by resonant modes of different Bloch wavenumbers. The
$Q$-factor of a nearby resonant mode  with Bloch wavenumber $\beta$ is
typically $O( 1/|\beta - \beta_*|^2)$. 
It has been shown that if the BIC  satisfies a special  condition,
then the $Q$-factor is at least $O(1/|\beta - \beta_*|^4)$ \cite{yuan18}. 
That condition was derived for BICs with or without symmetry
protection, and it was automatically satisfied by symmetric standing
waves which are BICs without symmetry protection \cite{yuan17_2}.  
In this section, we study ASWs satisfying that 
special condition, and show that the $Q$-factors of the nearby
resonant modes are typically 
$O(1/\beta^{6})$. 

As in the previous section, we assume the periodic structure given by
a dielectric function $\epsilon_*$ has a non-degenerate ASW $u_*$, where
$\epsilon_*$ is even in $y$,  $u_*$ is odd in $y$, and the frequency
$\omega_*$ of the ASW satisfies condition (\ref{cutoff1}). For 
$\beta$ near $\beta_*=0$, the resonant mode $u$, given as a Bloch mode
in Eq.~(\ref{eq:bloch}) for a periodic function $\phi$, has a complex frequency
$\omega$.  Assuming $|\beta L|$ is small, 
we expand $\phi$ and $\omega$ in power series of $\beta$:
\begin{eqnarray}
\label{omg_exp2}
\omega &=& \omega_* + \omega_1 \beta + \omega_2 \beta^2 + \omega_3
           \beta^3 + \omega_4 \beta^4 + \ldots, \\
\label{phi_exp} 
\phi &=& u_{\ast}  + \phi_1 \beta + \phi_2 \beta^2 + 
         \phi_3 \beta^3 + \phi_4 \beta^4 + \ldots ,
\end{eqnarray}
where $\phi_j$ for $j \geq 1$ are periodic in $y$ and must satisfy proper
outgoing radiation  conditions as $|x| \to \infty$. 
In terms of $\phi$, the Helmholtz equation becomes 
\begin{equation}  
\label{eq:phi}
\frac{\partial^2 \phi}{\partial x^2} + \frac{\partial^2 \phi}{\partial 
  y^2}  + 2 i \beta \frac{\partial \phi}{\partial y} + \left( k^2 
  \epsilon_* - \beta^2 \right) \phi = 0. 
\end{equation} 
 Inserting Eqs.~(\ref{omg_exp2})-(\ref{phi_exp}) into 
Eq.~(\ref{eq:phi}),  and comparing terms of equal power in $\delta$, we obtain 
\begin{eqnarray}
  \label{eq:phistar}
&&  {\cal L} u_* = 0, \\
\label{eq:phione}
&& {\cal L} \phi_1 =
 - 2i \partial_y u_* - 2 k_* k_1 \epsilon_*  u_*, 
\end{eqnarray}
where $\mathcal{L}$ is the operator defined in
Eq.~(\ref{defL}),  and  $k_j = \omega_j / c$ for $j \geq 1$. The equations for $\phi_2$,
$\phi_3$ and $\phi_4$ are also needed, and they will be given 
in simplified forms later in this section. 

We can determine $k_1$ such that
Eq.~(\ref{eq:phione}) has a solution. The result is
\begin{equation}
  \label{eq:k1_beta}
k_1 = \frac{\omega_1}{c} = \frac{ -  i 
  \int_\Omega \overline{u}_* \partial_y u_* \, d{\bm r} }
{k_{\ast} \int_\Omega \epsilon_* |u_*|^2 \, d{\bm r} } = 0. 
\end{equation}
Notice that $\overline{u}_* \partial_y u_*$ is odd in $y$, and thus
its integral on $\Omega$ is zero. Since
$\omega_*$ satisfies condition (\ref{cutoff1}) and the medium for 
$|x| > D$ is homogeneous, $\phi_1$ has the
asymptotic formulae
\begin{equation}
\label{phi1_exp}
\phi_1 \sim  d_0^{\pm} e^{\pm i k_{\ast  } x}, \quad x \to \pm
\infty, 
\end{equation}
where $d_0^{\pm}$ are unknown coefficients. Meanwhile, since $k_1=0$,
the equation for $\phi_2$ can be written as 
\begin{eqnarray}
\label{eq:phitwo}
  {\cal L} \phi_2 = - 2i \partial_y \phi_1    + (1 -2 k_* k_2 \epsilon_*) u_*.
\end{eqnarray}
The solvability of Eq.~(\ref{eq:phitwo}) allows us to determine
$k_2$. Following a procedure similar to that given in Appendix A, we obtain 
\begin{equation}
  \label{eq:k2_imag_beta}
  \mbox{Im}(k_2) 
  = \frac{\mbox{Im}(\omega_2)}{c} 
= - \frac{L   \left( |d_{0}^+|^2 +|d_{0}^-|^2 \right)} {2 
  \int_\Omega \epsilon_* |u_*|^2 \, d{\bm r}}.
\end{equation}
Therefore, the $Q$-factors in general satisfy $Q \sim \beta^{-2}$.
Multiplying the diffraction solutions $v_1$ and $v_2$ to
Eq.~(\ref{eq:phione}) and integrating on $\Omega$, we obtain
\begin{eqnarray}
   K_m =  i k_* L \left( d_0^-  \overline{S}_{1m} + d_0^+   \overline{S}_{2m} \right),
\end{eqnarray}
where 
\begin{equation}
  K_m = - i\int_\Omega \overline{v}_m  \partial_y u_* d {\bm r}, \quad
  m =1, 2.
\end{equation}
Clearly, the condition $d_0^- = d_0^+ = 0$ is equivalent to
$K_1 = K_2 = 0$.  

If $K_1 = K_2 = 0$, then $d_0^{\pm} = 0$ and $\phi_1 $ decays to
zero exponentially as $|x| \to \infty$. Since $k_1 = 0$, the
right hand side of Eq.~(\ref{eq:phione}) is even in $y$,
thus $\phi_1$ can be chosen as an even function of $y$. The right hand
side of Eq.~(\ref{eq:phitwo}) is odd in $y$, thus 
$\phi_2$ can be chosen as an odd function of $y$. 
Since there is only one opening  diffraction channel,  $\phi_2$ has
the asymptotic formulae
\begin{equation}
\label{phi2_exp}
\phi_2 \sim  d_1^{\pm} e^{\pm i k_{\ast  } x}, \quad x \to \pm
\infty
\end{equation}
for unknown coefficients $d_1^{\pm}$. However, as in the previous
section, due  to the symmetry mismatch between $\phi_2$ and plane
waves $e^{\pm i k_* x}$, we can show that $d_1^- = d^+_1 = 0$. 

With $k_1=0$, the equation for $\phi_3$ becomes 
\begin{equation}
\label{eq:phi3} 
\mathcal{L} \phi_3 = - 2 i \partial_y \phi_2 + (1 - 2 k_* k_2 \epsilon_*) \phi_1 - 2 k_* k_3 \epsilon_* \phi_*.
\end{equation}
It is easy to show that the solvability of $\phi_3$ gives rise to $k_3 = 0$. 
Since $k_1=k_3=0$, the equation for $\phi_4$ becomes
\begin{equation*}
\mathcal{L} \phi_4 = - 2 i \partial_y \phi_3 + (1 - 2 k_* k_2 \epsilon_*) \phi_2 - (2 k_* k_4 + k_2^2) \epsilon_* \phi_*.
\end{equation*}
Following a procedure similar to that in Appendix C,   we can show that 
\begin{equation}
  \label{eq:k4_imag_beta}
  \mbox{Im}(k_4) 
  = \frac{\mbox{Im}(\omega_4)}{c} 
= - \frac{L   \left( |d_{1}^+|^2 +|d_{1}^-|^2 \right)} {2 
  \int_\Omega \epsilon_* |u_*|^2 \, d{\bm r}} = 0.
\end{equation}
Furthermore, $\mbox{Im}(\omega_5) $ must be zero, since otherwise 
$\mbox{Im}(\omega)$ can have the wrong sign when $\beta$ changes 
signs.  Note that the right hand side of Eq.~(\ref{eq:phi3}) is an
even function of $y$, therefore, $\phi_3$ may radiate out power as $x \to
\pm \infty$, we expect $\mbox{Im}(\omega_6)$ to be nonzero, 
$\mbox{Im}(\omega) \sim \beta^6$ and $Q \sim \beta^{-6}$. 

\section{Numerical examples}
\label{sec:example}

\begin{figure}[htb]
\centering 
\includegraphics[scale=0.6]{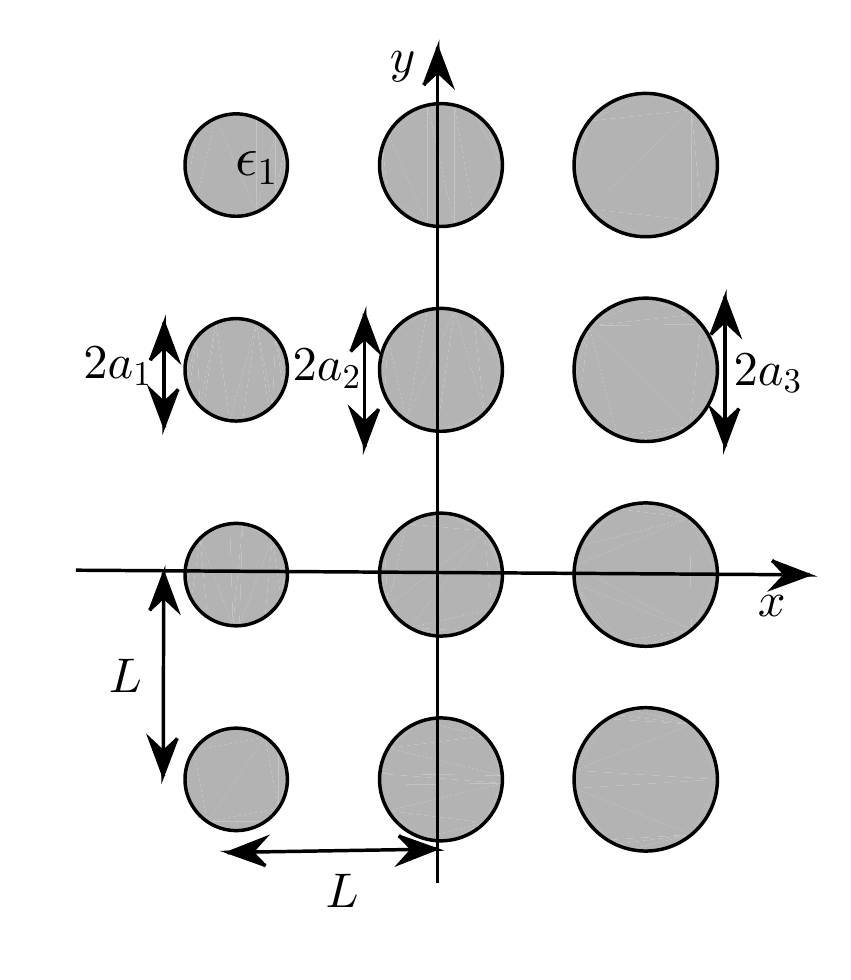} 
\caption{A periodic structure (periodic in $y$ with period $L$) with
  three arrays of circular dielectric cylinders.}
\label{figone}
\end{figure}

To  validate and illustrate the theoretical results developed in the
previous sections, we present numerical examples for periodic
structures consisting of three arrays of circular dielectric
cylinders surrounded by air. 
As shown  in Fig.~\ref{figone}, 
the arrays are periodic  in the $y$ direction with a period $L$ and
separated by a center-to-center distance $L$ in the $x$ direction. 
The radii of the cylinders in the three arrays are $a_1$, $a_2$ and
$a_3$, respectively, and the dielectric constants of all cylinders are
$\epsilon_1$. The coordinates are chosen so that  the center of one
cylinder is at the origin and the centers of the cylinders of the
second array are on the $y$ axis. The domain $\Omega$ contains three 
cylinders. 

\begin{figure}[htb]
\centering 
\includegraphics[scale=0.55]{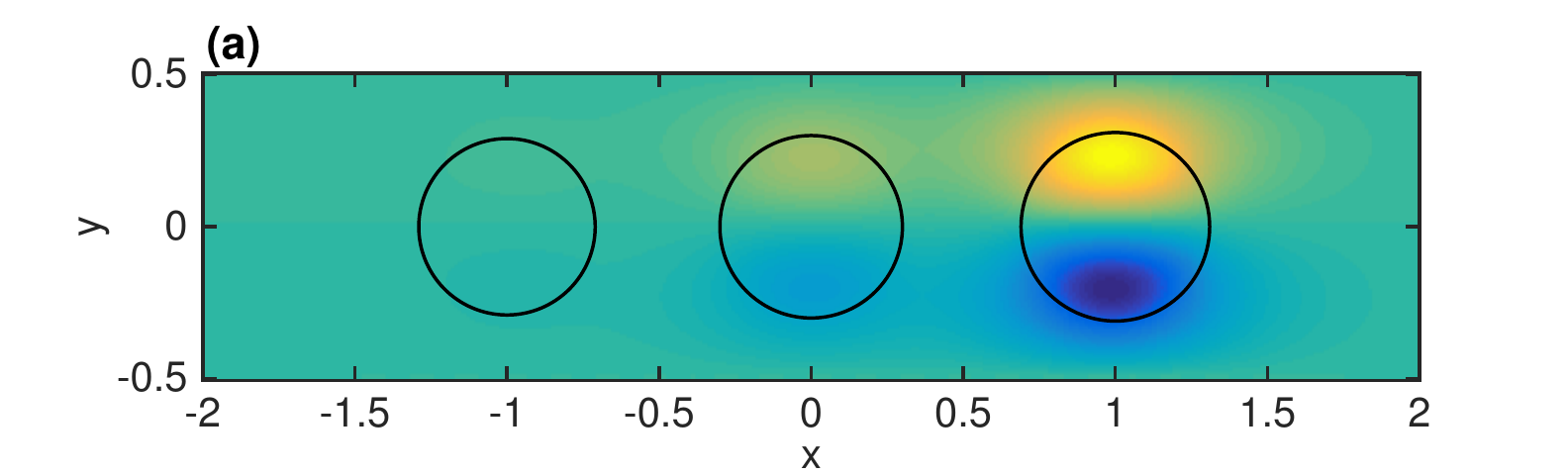} 
\includegraphics[scale=0.55]{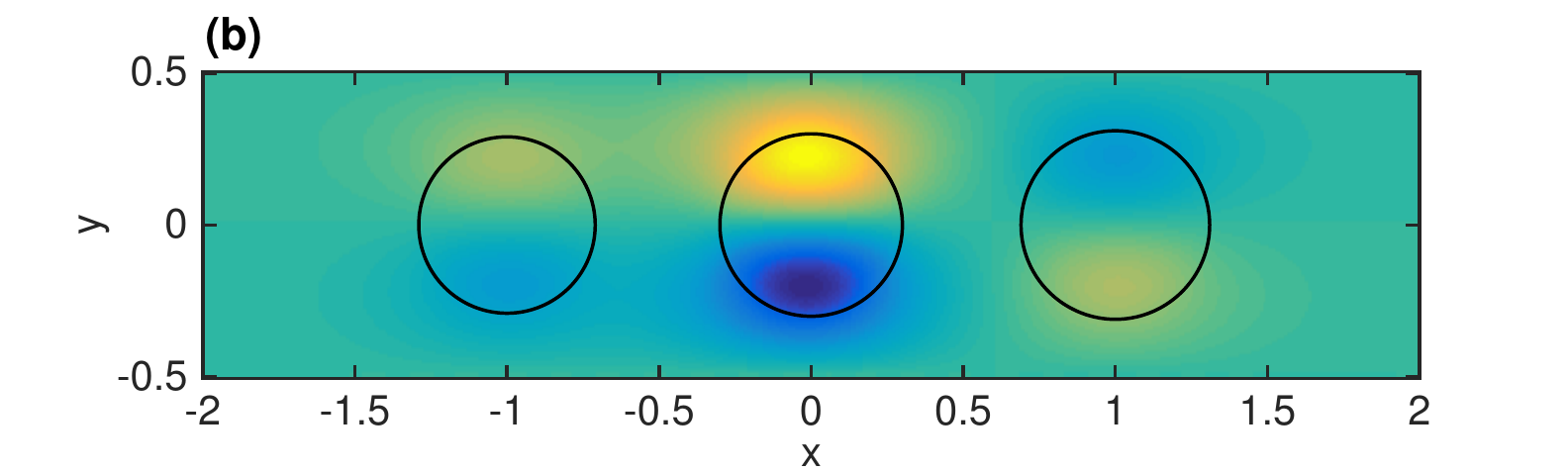} 
\includegraphics[scale=0.55]{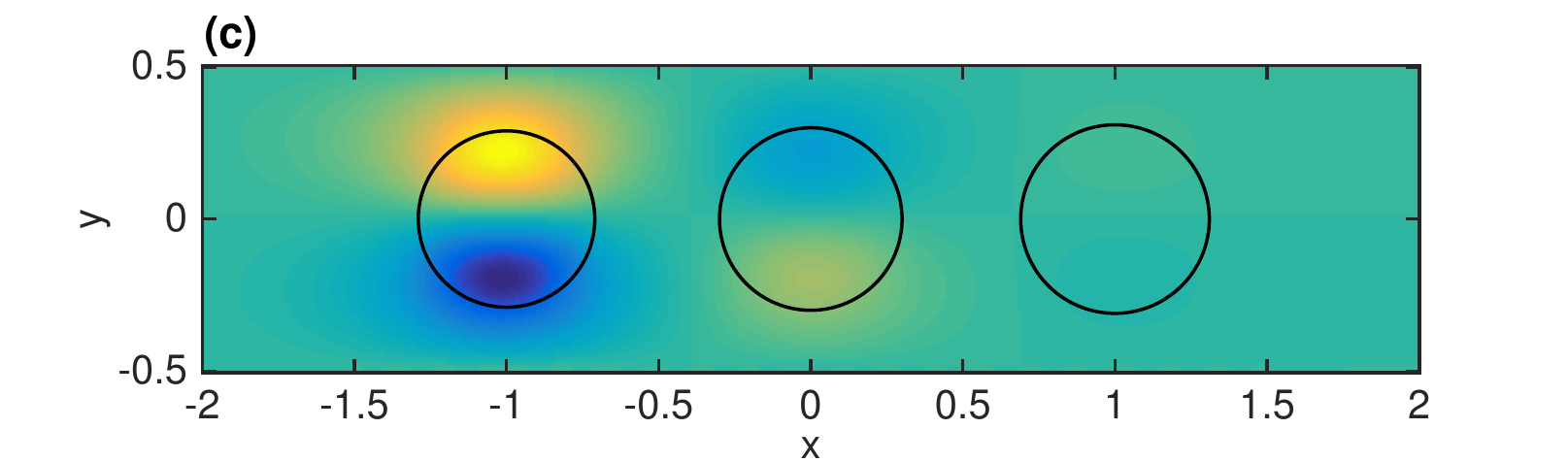} 
\caption{Wave field patterns of antisymmetric standing waves 
  on a periodic structure with three arrays of cylinders: (a) ASW1,
  (b) ASW2, (c) ASW3.}
\label{figtwo}
\end{figure}

\begin{figure}[htb]
\centering 
\includegraphics[scale=0.6]{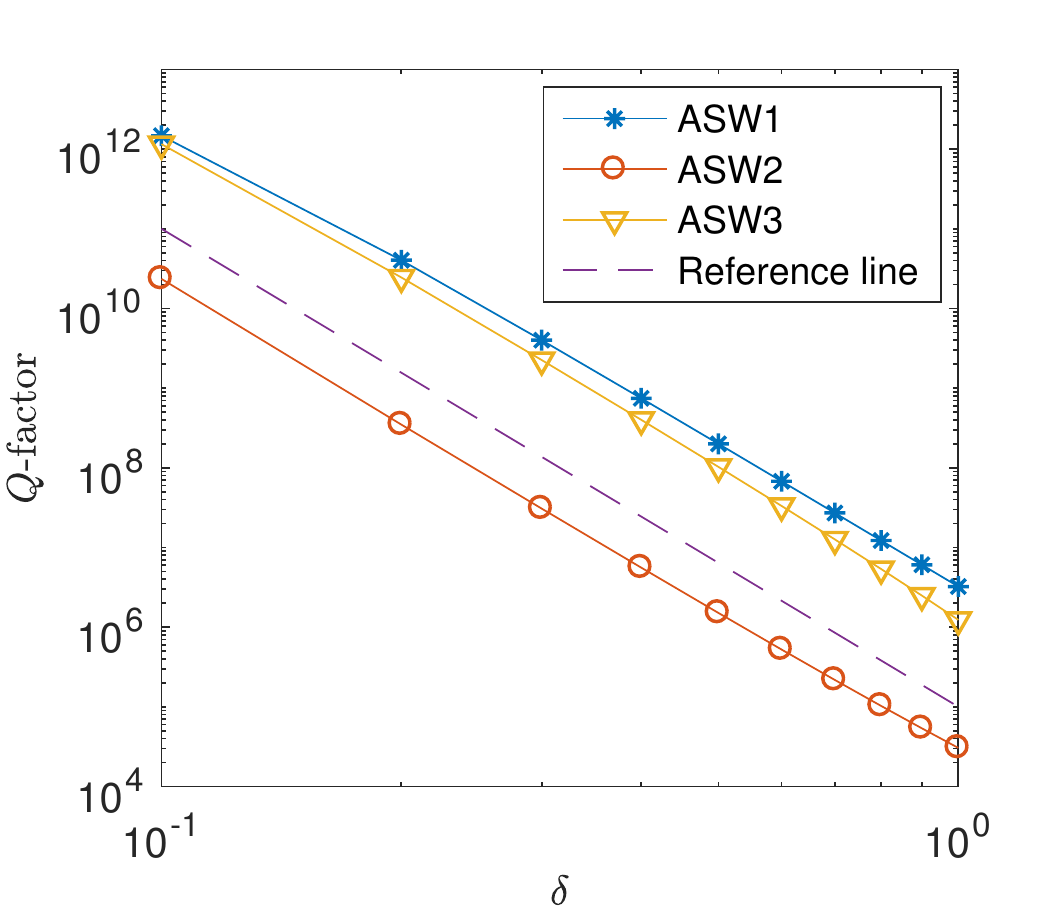} 
\caption{Example 1: $Q$-factors of resonant modes near ASW1, ASW2 and
  ASW3 for perturbed structures with coefficients of $F$ listed in
  Table~\ref{table1}. The dashed line is the reference line for $Q =
  10^{5} / \delta^6$.} 
\label{figthree}
\end{figure}

To consider structural perturbations, we let $a_1 = 0.29L$, $a_2 =
0.3L$,  $a_3 = 0.31$ and $\epsilon_1 = 4$. 
For the $E$ polarization, 
the structure has three ASWs with normalized frequencies 
$\omega_* L / (2\pi c) = 0.6575$, $0.6715$ and $0.6862$,
respectively, and they are referred to as ASW1, ASW2 and ASW3. 
The electric field patterns of these three ASWs shown in
Fig.~\ref{figtwo}. 
For the  first example, we consider a perturbation profile $F$ given by
\begin{equation}
\label{defFF} F(x,y) =  \sum_{j=1}^3 \eta_j F_j(x,y),
\end{equation}
where $\eta_1$, $\eta_2$ and $\eta_3$ are real constants, $F_1$, $F_2$
and $F_3$ are defined (on domain $\Omega$) by 
\begin{equation} 
\label{defF1_odd} F_j(x,y) = \left\{  
\begin{array}{ll} \sin \dfrac{\pi y}{2 a_j}, &
\mbox{in the $j$-th cylinder}, \\
 0, &  \mbox{ otherwise},  \end{array} \right. 
\end{equation}
for $j=1$, 2 and 3. We are interested in a profile $F$ such that 
$G_1 =  G_2 = 0$,  i.e. 
\[
\sum_{j=1}^3 \eta_j \int_\Omega F_j u_* \overline{v}_m \, d {\bm r}  =
0, \quad m=1, 2.
\]
As we mentioned in Sec.~\ref{sec:material_pert}, the condition
$G_1=G_2=0$ is equivalent to two real independent conditions,
thus, by fixing any one of $\eta_1$, $\eta_2$ and $\eta_3$, we
can solve the other two from the above system.  After that, 
we can scale $F$ such that $\mbox{max}_{(x,y) \in \Omega} |F| =
1$. 
The obtained values of  $\eta_1$, $\eta_2$ and $\eta_3$, 
corresponding to the three ASWs on the unperturbed structures, are
listed in Table~\ref{table1} 
\begin{table}[htp]
\centering 
\begin{tabular}{c|c||c|c|c} \hline \hline
BICs & $\omega_* L/(2\pi c)$  &  $\eta_1$ & $\eta_2$ & $\eta_3$ \\ \hline 
ASW1 & $0.6575$  & $1$ & $-0.5077$  & $0.1255$\\ \hline
ASW2 & $0.6715$ & $1$ &   $-0.2380$   & $0.2445$   \\ \hline
ASW3 & $0.6862$ & $0.0844$ & $ 0.4679$ & $1$ \\ \hline  \hline
\end{tabular}
\caption{Example 1: coefficients of antisymmetric perturbation
  profiles $F$ satisfying the
  condition $G_1=G_2=0$.}
\label{table1} 
\end{table}
below. 
Since $F_1$, $F_2$ and $F_3$ are odd in $y$, $F$ is also odd in $y$,
thus the $Q$-factors of the resonant modes should satisfy $Q \sim
\delta^{-6}$. In Fig.~\ref{figthree}, 
we show the $Q$-factors of the resonant modes near the three ASWs for
different values of $\delta$.  The numerical results clearly confirm
the inverse sixth order relation between $Q$ and $\delta$. 

To calculate the $Q$-factors of the resonant modes, we solve an
eigenvalue problem formulated using 
a mixed Fourier-Chebyshev pseudospectral method to discretize the
three circular disks  (corresponding to the cross
section of the three cylinders in $\Omega$) \cite{tref00}, and
cylindrical and plane wave expansions outside the cylinders
\cite{huang06}.   

For the second example, we consider perturbation profiles 
that are asymmetric  in the $y$ direction. For the same unperturbed
structure with three arrays of cylinders and the same $F_2$ and $F_3$
given in Eq.~(\ref{defF1_odd}), we define  a new $F_1$ (on domain $\Omega$)  by
\begin{equation*} 
 F_1(x,y) = \left\{  \begin{array}{ll} \sin \left( \dfrac{  \pi y}{2a_1} + \dfrac{\pi}{4} \right), &  \mbox{ in the first cylinder}, \\
                                                         0, & \mbox{ otherwise},  \end{array} \right. 
\end{equation*}
For $F$ given in Eq.~(\ref{defFF}), we solve 
$\eta_1$, $\eta_2$ and $\eta_3$ such that $G_1 = G_2 = 0$ 
for each ASW. The results are listed in Table~\ref{table2}
\begin{table}[htp]
\centering 
\begin{tabular}{c|c||c|c|c} \hline \hline
BICs & $\omega_* L/(2\pi c)$  &  $\eta_1$ & $\eta_2$ & $\eta_3$ \\ \hline 
ASW1 & $0.6575$  & $1$ & $-0.3583$  & $0.0885$\\ \hline
ASW2 & $0.6715$ & $1$ &   $-0.1683$   & $0.1729$   \\ \hline
ASW3 & $0.6862$ & $0.1194$ & $0.4679$ & $1$ \\ \hline \hline
\end{tabular}
\caption{Example 2: coefficients of asymmetric perturbation profiles
  satisfying the condition $G_1= G_2 = 0$.}
\label{table2} 
\end{table}
below. Since $F$ is neither even in $y$ nor odd in $y$, 
the $Q$-factors of nearby resonant modes should satisfy $Q \sim \delta^{-4}$. 
Numerical results for the $Q$-factors are shown in Fig.~\ref{figthree}
\begin{figure}[htb]
\centering 
\includegraphics[scale=0.58]{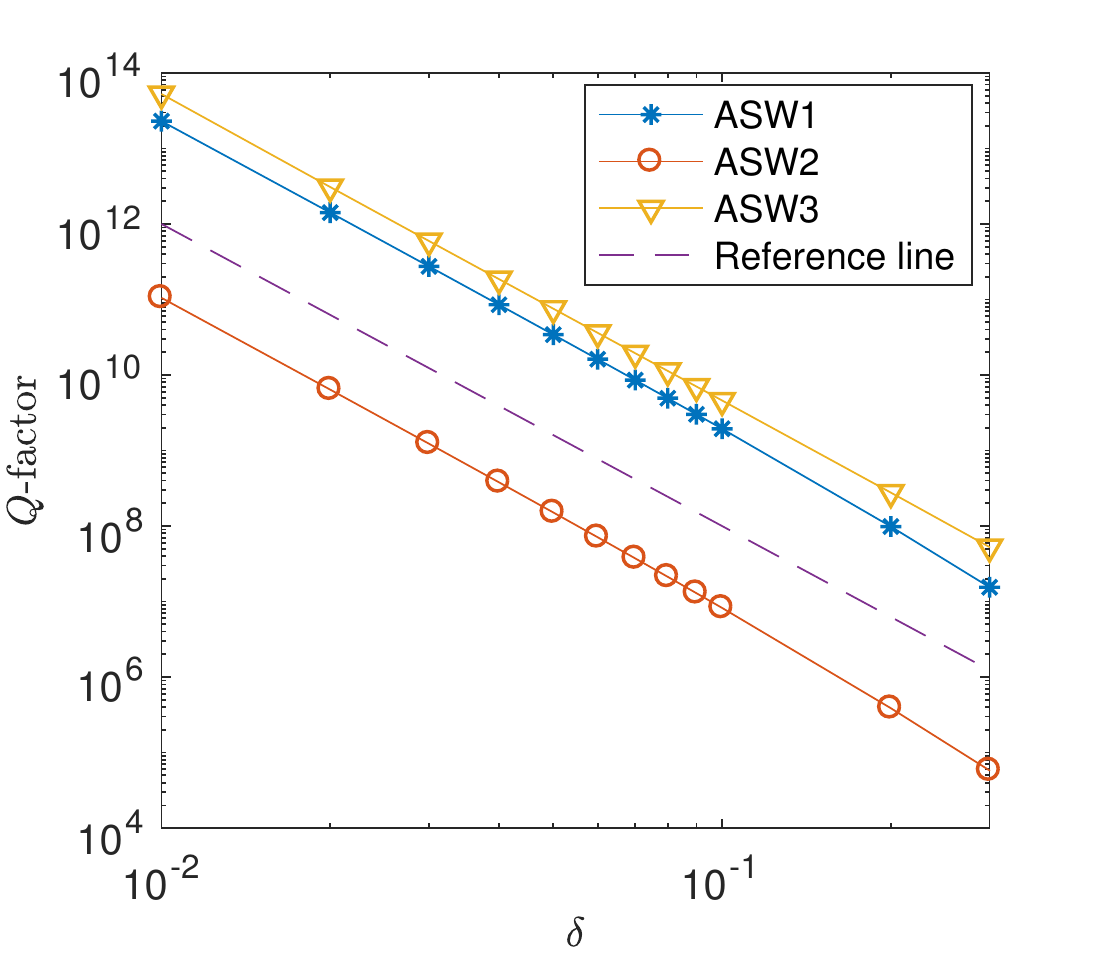} 
\caption{Example 2: $Q$-factors of resonant modes near ASW1, ASW2 and
  ASW3 for perturbed structures with coefficients of $F$ 
  listed in Table~\ref{table2}. The dashed line is the
  reference line for $Q = 10^{4} / \delta^4$.} 
\label{figfour}
\end{figure}
in logarithmic scales, and they agree very well with the inverse
fourth order relation between $Q$ and $\delta$.

\begin{figure}[htb]
\centering 
\includegraphics[scale=0.6]{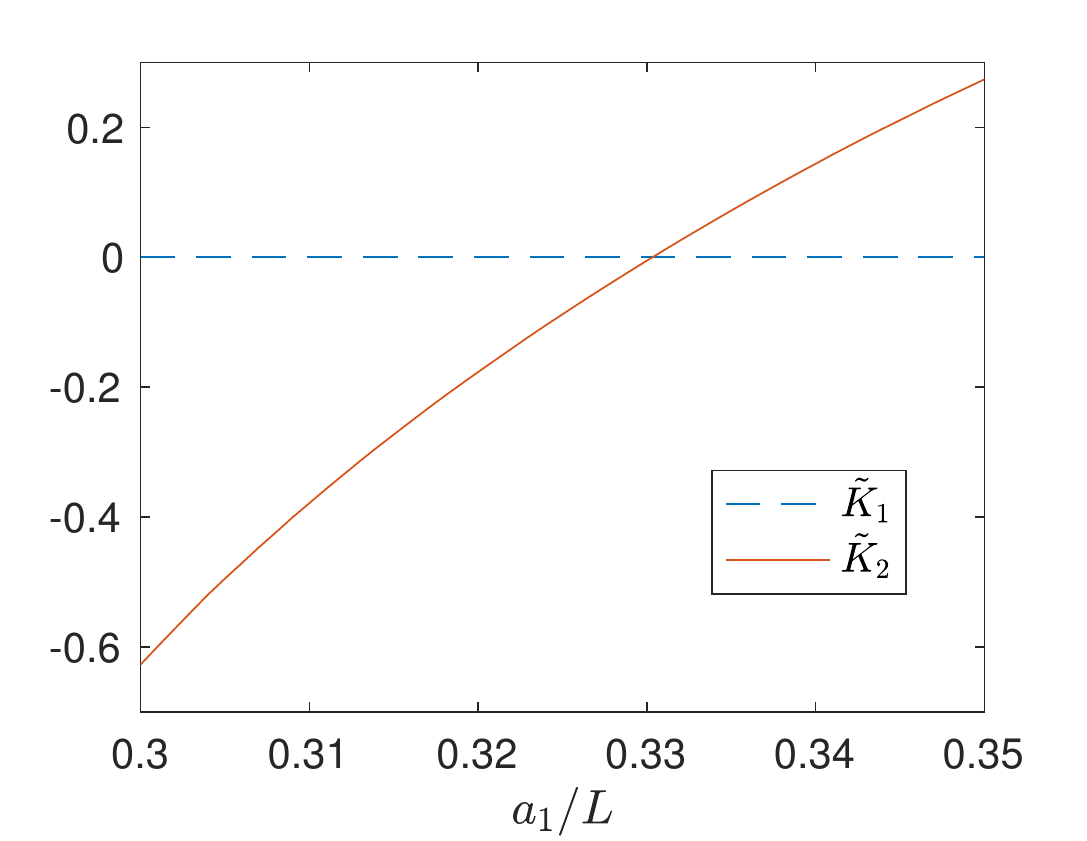} 
\caption{Example 3: $\tilde{K}_1$ and $\tilde{K}_2$ [defined in
  Eq.~(\ref{Ktilde}), with unit $L$] as functions of $a_1$ for an ASW 
  on a periodic structure with three arrays of cylinders with radii
  $a_1$, $a_2 = 0.3L$ and $a_3=a_1$.}
\label{figfive}
\end{figure}

\begin{figure}[htb]
\centering 
\includegraphics[scale=0.55]{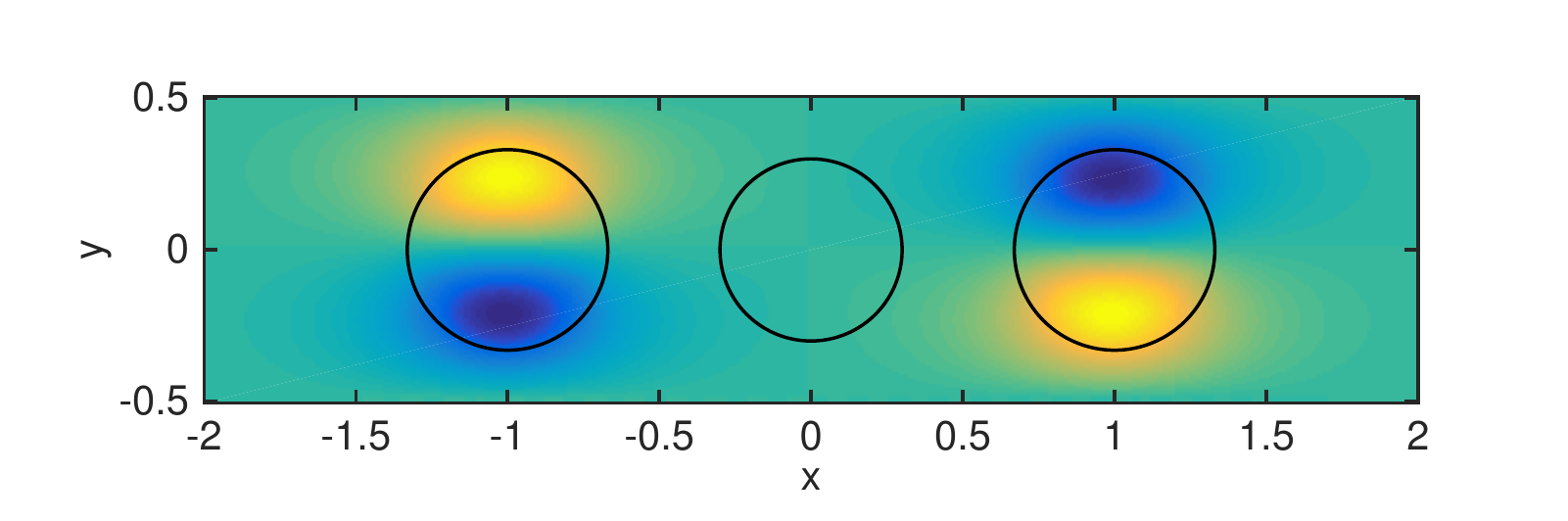} 
\caption{Example 3: Wave field pattern of an ASW with frequency 
$\omega_* L/(2 \pi c) = 0.6361$ on a periodic
  structure with $a_1= a_3 = 0.33033 L$ and $a_2 = 0.3 L$.}
\label{figsix}
\end{figure}

The third example is designed to validate  the
perturbation results of Sec.~\ref{sec:beta_pert}.  
 We consider the same periodic structure with three arrays of
 cylinders as shown in  Fig.~\ref{figone}, keep $a_2 = 0.3L$ and
 $\epsilon_1=4$ as in the first example, but assume $a_1=a_3$. 
First, we show that for $a_1 = a_3 \approx 0.33033L$, the periodic structure
has an ASW satisfying the condition $K_1=K_2 = 0$.  As shown in
Appendix D, the diffraction solutions $v_1$ and $v_2$ have two linear
combinations $w_1$ and $w_2$ that are real even
functions of $y$. Since the periodic structure of this example is
symmetric in $x$,  $w_1$ and $w_2$  can be
be written down explicitly as 
\[
w_1 = \frac{ e^{-i \theta_1}}{\sqrt{2}} (v_1 + v_2), 
\quad 
w_2 = \frac{ e^{-i \theta_2}}{\sqrt{2}} (v_1 - v_2), 
\]
where $\theta_1$ and $\theta_2$ satisfy 
$R+T = e^{ 2 i \theta_1}$ and $R- T = e^{ 2 i \theta_2}$, 
 $R=S_{11}=S_{22}$ and $T=S_{12}=S_{21}$ are entries of 
the scattering matrix ${\bm S}$. 
Importantly,  $w_1$ is even in $x$ and $w_2$ is  odd in $x$.
Clearly, the condition $K_1=K_2 = 0$ is equivalent to 
$\tilde{K}_1=\tilde{K}_2 = 0$, where 
\begin{equation}
  \label{Ktilde}
\tilde{K}_m = \int_\Omega \overline{w}_m \partial_y u_* d{\bm r},
\quad m=1, 2.  
\end{equation}
In Fig.~{\ref{figfive}}, 
we show $\tilde{K}_1$ and $\tilde{K}_2$ as
functions of $a_1$ for an ASW with a wave field pattern similar to
that of ASW2 in Fig.~\ref{figtwo}(b). The ASW is normalized such that 
$\int_\Omega |u_*|^2 d{\bm r} /L^2  = 1$. 
 Since the ASW is odd in $x$, 
we have $\tilde{K}_1=0$. From Fig.~\ref{figfive},  it is clear  that
$\tilde{K}_2$ is real and changes signs. Therefore, $\tilde{K}_2$ must have a zero. Solving 
$\tilde{K}_2 = 0$ by a numerical method,  we obtain $a_1 = a_3 = 0.33033L$. 
The normalized frequency of the corresponding ASW is
$\omega_* L/(2 \pi c) = 0.6361$. Its wave field pattern is shown in
Fig.~{\ref{figsix}}. 

\begin{figure}[htb]
\centering 
\includegraphics[scale=0.55]{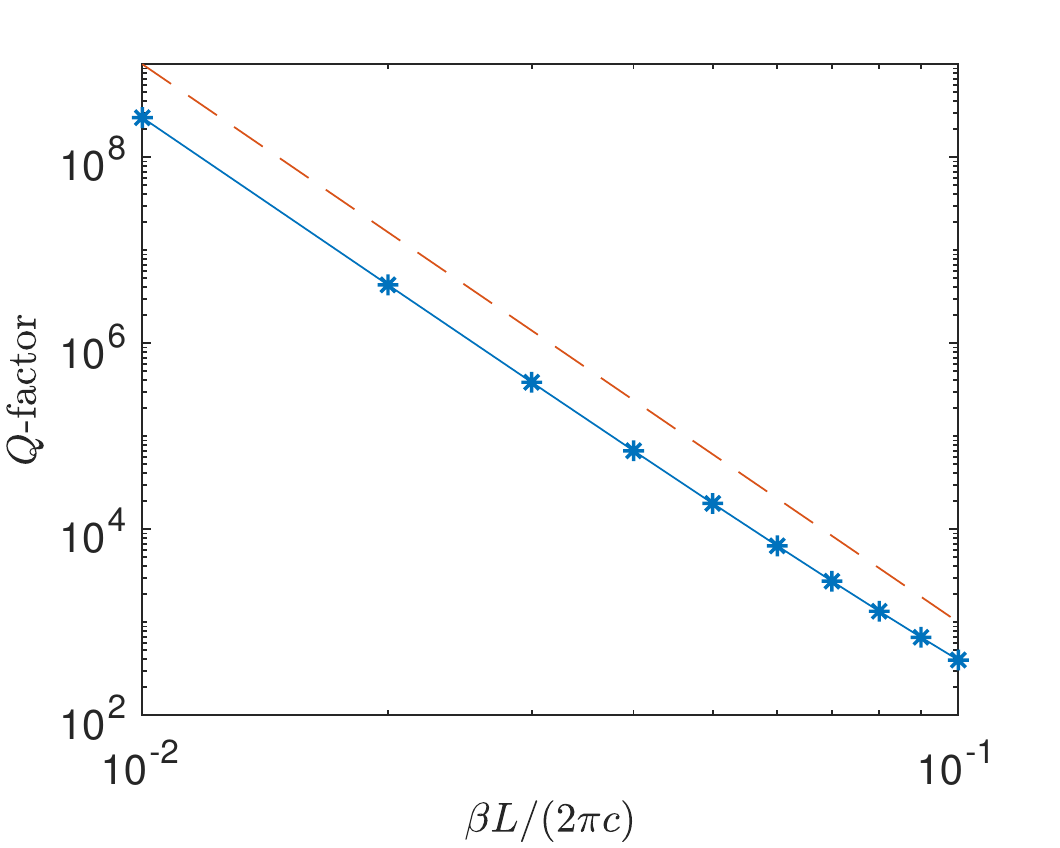} 
\caption{Example 3: $Q$-factors of resonant modes near the ASW of
  Fig.~\ref{figsix}. The dashed line is the reference line for $Q =
  10^{-3} [ 2\pi /(\beta L)]^6$. } 
\label{figseven}
\end{figure}

Finally, we calculate the $Q$-factors of  the 
nearby resonant modes with $\beta$ close to 0. 
The numerical results are shown in Fig.~{\ref{figseven}}, 
and they agree well with the inverse sixth order relation between $Q$
and $\beta$.  

\section{Conclusion}
\label{conclusion}

The BICs are useful mainly because high-$Q$ resonances can be easily
created by slightly changing the structure or other parameters. For
theoretical understanding and practical applications, it 
is important to find out how the $Q$-factors of the resonant modes
depend on the amplitude of the perturbations. Near a
symmetry-protected BIC, a symmetry-breaking structural perturbation of
amplitude $\delta$ typically creates a resonant mode with a $Q$-factor
proportional to $1/\delta^2$. For ASWs on 2D periodic structures with a
reflection symmetry, we show  that the $Q$-factor can be
$O(1/\delta^4)$ if the perturbation satisfies a simple condition, and
it becomes $O(1/\delta^6)$ if in addition to that condition, the
perturbation is antisymmetric.  On the other hand, a BIC on a periodic
structure is always
surrounded by resonant modes depending on the Bloch
wave vector. It is well known that the $Q$-factors of these resonant
modes tend to infinity as the wave vector tends to that of the BIC. 
For 2D periodic structures, we show that if an ASW satisfies a
condition first derived in \cite{yuan18},  the $Q$-factors of nearby 
resonant modes are $O(1 / \beta^6)$, where $\beta$ is the Bloch
wavenumber. These theoretical results can be useful for practical
applications where high-$Q$ resonances are needed. The $O(1/\delta^6)$
result on structural perturbation implies that large $Q$-factors are
possible without using a very small $\delta$, probably making the
fabrication process somewhat easier. The $O(1/\beta^6)$ result on
wavenumber perturbation implies that strong resonances can be induced
by incident waves with a wider range of incident angles. 

The current study is limited to ASWs which are symmetry protected
BICs. For BICs unprotected by symmetry, we have derived a condition on the 
perturbation profile $F$ such that the $Q$-factors of nearby resonant
modes are proportional to $1/\delta^4$, but have not found a condition for 
$Q \sim 1/\delta^6$.   Our study is also limited to 2D structures with
a 1D periodicity. It is worthwhile to find similar results for 
three-dimensional biperiodic structures such as photonic crystal
slabs. On actual fabricated structures, the $Q$-factors are
limited by many practical issues, such as fabrication errors, material
losses, finite sizes, and loss of periodicity (for periodic
structures). It is necessary to have a comprehensive study on the
different factors, but we still expect the structures satisfying the
conditions for high-$Q$ resonances have advantages in realistic
devices. 

\section*{Acknowledgments}
The authors acknowledge support from the Natural Science Foundation of Chongqing, China (Grant No. cstc2019jcyj-msxmX0717),  the Science and Technology Research Program of Chongqing
Municipal Education Commission, China (Grant No. KJ1706155),  and the Research Grants
Council of Hong Kong Special Administrative Region, China (Grant
No. CityU 11304117).


\section*{Appendix A}
It is easy to verify that 
$\int_\Omega \overline{u}_* {\cal L} u_2 d{\bm r} = 0$. 
See Appendix in Ref.~\cite{yuan18}. Multiplying both sides of Eq.~(\ref{eq:utwo}) and integrating on
$\Omega$, we obtain
\[
k_2 =  \frac{  \int_\Omega R\, d{\bm r}  -  \int_\Omega (k_1^2 \epsilon_* + 2 k_* k_1 F) |u_*|^2 d{\bm r} 
 }{ 2 k_* \int_\Omega \epsilon_* |u_*|^2 
  d{\bm r}},
\]
where $R = - (2 k_* k_1 \epsilon_* + k_*^2 F) \overline{u}_* u_1 $. Therefore, 
\[
\mbox{Im}(k_2) =  \frac{  \mbox{Im} \left[ \int_\Omega R\, d{\bm r} \right] }{ 2 k_*
  \int_\Omega \epsilon |u_*|^2    d{\bm r}}.
\]
From the complex conjugate of Eq.~(\ref{eq:uone}), we obtain
\[
\int_{\Omega_h} 
u_1 \overline{\cal L} \overline{u}_1  d{\bm r} 
= \int_{\Omega_h} R\, d{\bm r},
\]
where $\Omega_h$ is a rectangle given by $|y| < L/2$ and $|x| < h$ for $h > D$. It is easy to verify that 
\begin{eqnarray*}
&& \int_{\Omega_h} u_1 \overline{\cal L} \overline{u}_1  d{\bm r} 
= 
\int_{\partial \Omega_h} 
u_1 \frac{\partial \overline{u}_1}{\partial \nu} ds  \cr
&& + \int_{\Omega_h} \left( k_*^2 \epsilon_*  |u_1|^2  -
   |\nabla u_1|^2   \right) d{\bm r}, 
\end{eqnarray*}
where $\partial \Omega_h$ is the boundary of $\Omega_h$ and $\nu$ is
its unit outward normal vector. The second term in the right hand side
above is real. Therefore, 
\[
\mbox{Im} \left[ \int_{\Omega_h} R\, d{\bm r} \right] 
=  \mbox{Im} \left[ 
\int_{\partial \Omega_h} 
u_1 \frac{\partial \overline{u}_1}{\partial \nu} ds \right].
\]
Since $u_1$ is periodic in $y$, the line integrals at $y=\pm L/2$ are
canceled. Therefore
\[
\int_{\partial \Omega_h} 
u_1 \frac{\partial \overline{u}_1}{\partial \nu} ds
= 
\int_{-L/2}^{L/2} \left[ 
 u_1 \frac{\partial \overline{u}_1}{\partial x}
\right]^{x=h}_{x=-h} dy,
\]
where $P(x,y)|_{x=-h}^{x=h}$ denotes $P(h,y)-P(-h,y)$. 

For $|x| > D$, the equation for $u_1$ is quite simple. It is not
difficult to see that 
\[
u_1 = b_0^\pm e^{ \pm i k_* x} 
+ \sum_{j\ne 0} f_j^\pm(x) e^{ i 2\pi jy/L} e^{\pm \gamma_{*j} x}
\]
for $x> D$ and $x< -D$ respectively, where $\gamma_{*j} = -i
\sqrt{k_*^2 - (2 j \pi/L )^2}$ for $j \neq 0$ is positive, 
$b_0^\pm$ are unknown coefficients, and $f_j^\pm(x)$ ($j\ne 0$) are unknown
linear polynomials of $x$. The above gives
\[
\lim_{h \to +\infty} \int_{-L/2}^{L/2} \left[ 
 u_1 \frac{\partial \overline{u}_1}{\partial x}
\right]^{x=h}_{x=-h} dy 
= -i k_* L (|b_0^+|^2 + |b_0^-|^2), 
\]
and $\mbox{Im} \left[ \int_\Omega R\, d{\bm r} \right] = - L k_*
(|b_0^+|^2 +   |b_0^-|^2)$, and finally Eq.~(\ref{eq:k2_imag}).

\section*{Appendix B}
To show Eq.~(\ref{leftinc}), we notice that 
\[
\overline{v}_1 {\cal L} u_1 
-u_1 \overline{\cal L} \overline{v}_1 = 
\nabla \cdot [ \overline{v}_1 \nabla u_1 
- u_1 \nabla \overline{v}_1 ].
\]
Since $v_1$ satisfies the Helmholtz equation and both $u_1$
and $v_1$ are periodic in $y$, we have 
\[
\int_{\Omega_h} \overline{v}_1 {\cal L} u_1 d{\bm r}
= 
\int_{\partial \Omega_h} \left[ \overline{v}_1 \frac{ \partial
    u_1 }{\partial \nu} 
- u_1 \frac{ \partial \overline{v}_1 }{\partial \nu} \right] ds. 
\]
In the right hand side above, the integrals on the two edges at
$y=\pm L/2$ are canceled. Therefore 
\[
\int_{\Omega_h} \overline{v}_1 {\cal L} u_1 d{\bm r}
= 
\int_{-L/2}^{L/2}  \left[ \overline{v}_1 \frac{ \partial
    u_1}{\partial x}
- u_1 \frac{ \partial \overline{v}_1 }{\partial x} \right]^{x=h}_{x=-h} dy.
\]
Based on the asymptotic formula (\ref{leftasym}), it is easy to show
that as $h \to +\infty$, the right hand side above tends to $2i k_* L
(b_0^- \overline{S}_{11}   + b_0^+ \overline{S}_{21})$. This leads to
Eq.~(\ref{leftinc}) for $m=1$. The case for $m=2$ is similar.

\section*{Appendix C}
Inserting Eqs.~(\ref{u_exp})-(\ref{omg_exp}) into 
Eq.~(\ref{eq:helm}),  and comparing terms of equal power in 
$\delta$, we obtain the following equation for $u_3$
\begin{eqnarray}
\nonumber
  {\cal L} u_3 & = & -   k_*^2 F u_2   - 2 k_* k_2 \epsilon_*  u_1  \\ 
  \label{eq:uthree}
& &  - (2 k_* k_3 \epsilon_* +   2 k_* k_2 F ) u_*  
\end{eqnarray}
In the above,  the result $k_1 = 0$ is used.  Multiplying both sides
of Eq.~(\ref{eq:uthree}) by $\overline{u}_*$ and integrating on
$\Omega$, we have 
$$ k_3 = - \frac{\int_\Omega k_* F u_2 \bar{u}_* + 2 k_2 \epsilon_* u_1 \bar{u}_* + 2 k_2 F |u_*|^2 d {\bf r}}{2 \int_\Omega \epsilon_* |u_*|^2 d {\bf r}} .$$
Note that $F$, $u_*$ and $u_2$ are odd functions of $y$, and
$\epsilon_*$ and $u_1$ are even functions of $y$, thus $k_3 = 0$.

The governing equation for $u_4$ is
\begin{eqnarray}
\nonumber
 {\cal L} u_4 &=&  -  k_*^2 F u_3   - 2 k_* k_2 \epsilon_*  u_2  -  2 k_* k_2 F  u_1 \\
 \label{eq:ufour}
&& - (2 k_* k_4 \epsilon_* + k_2^2 \epsilon_* ) u_*. 
\end{eqnarray}
The results $k_1 = k_3 = 0$ have been used to simplify the above equation. 
Multiplying both sides of Eq.~(\ref{eq:ufour}) and integrating on
$\Omega$, we obtain
\[
k_4 =  \frac{  \int_\Omega W \, d{\bm r}  - k_2^2 \int_\Omega   \epsilon_*  |u_*|^2 d{\bm r} 
 }{ 2 k_* \int_\Omega \epsilon_* |u_*|^2 
  d{\bm r}},
\]
where 
$$W = -  2 k_* k_2 F  \overline{u}_* u_1 - 2 k_* k_2 \epsilon_* \overline{u}_* u_2  -  k_*^2 F \overline{u}_* u_3.$$ 
Note that $k_2$  is real. Therefore, 
\[
\mbox{Im}(k_4) =  \frac{  \mbox{Im} \left[ \int_\Omega W\, d{\bm r} \right] }{ 2 k_*
  \int_\Omega \epsilon |u_*|^2    d{\bm r}}.
\]

From Eqs.~(\ref{eq:uone}), (\ref{eq:utwo}) and (\ref{eq:uthree}), we have
$$ \mbox{Im} \left[   \int_{\Omega_h} ( {u}_1 \overline{\mathcal{L}} \overline{u}_3 - \overline{u}_3 \mathcal{L} u_1  +   {u}_2 \overline{\cal L} \overline{u}_2 ) d {\bm r} \right] = \mbox{Im} \left[ \int_{\Omega_h} W d {\bm r}  \right].$$
Since $u_1$ decays to zero exponentially, $\int_\Omega \overline{u}_3
\mathcal{L} u_1 d {\bm r}$ and $\int_\Omega {u}_1
\overline{\mathcal{L}} \overline{u}_3 d {\bm r}$ are well defined. It
is easy to show that 
$\int_\Omega ( {u}_1 \overline{\mathcal{L}} \overline{u}_3 - \overline{u}_3 \mathcal{L} u_1  )d {\bm r} = 0$.  
On the other hand, due to the asymptotic formula (\ref{u2_exp}), we have 
$$ \lim\limits_{h \to \infty}  \int_{\Omega_h} {u}_2 \overline{\cal L} \overline{u}_2 d {\bm r}  = - i  k_* L \left( | b_1^- |^2 + | b_1^+ |^2 \right). $$
Therefore, 
\begin{eqnarray*}
&&  \lim\limits_{h \to \infty } \int_{\Omega_h} ( {u}_1
   \overline{\mathcal{L}} \overline{u}_3 - \overline{u}_3 \mathcal{L}
   u_1  +   {u}_2 \overline{\cal L} \overline{u}_2 ) d {\bm r} \cr
&& = -i L
   k_*  \left( | b_1^- |^2 + | b_1^+ |^2 \right).
\end{eqnarray*} 
That means
\[
\mbox{Im} \left[ \int_{\Omega} W d {\bm r} \right] =  - L k_*  
\left( | b_1^- |^2 + | b_1^+ |^2 \right).
\]
This leads to Eq.~(\ref{eq:k4_imag}).

\section*{Appendix D}
The $2\times 2$ scattering matrix ${\bm S}$ is symmetric and unitary
due to reciprocity and conservation of energy. Therefore, it has two real
orthonormal eigenvectors ${\bm a}_1$ and ${\bm a}_2$. The
corresponding eigenvalues $\lambda_1$ and $\lambda_2$ have  unit
magnitudes, i.e., for $j=1$ and 2, $\lambda_j = \exp(  i 2\theta_j)$
where $\theta_j$ is real.  Writing down
the eigenvectors as 
\[
{\bf a}_1 = \left[ \begin{matrix}
    a_{11} \cr a_{21} \end{matrix} \right], \quad 
{\bf a}_2 = \left[ \begin{matrix}
    a_{12} \cr a_{22} \end{matrix} \right], 
\]
we define two functions $w_1$ and $w_2$ by 
\[
w_j = e^{-i \theta_j} ( a_{1j} v_1 + a_{2j} v_2), \quad j=1, 2.
\]
then 
\[
w_j \sim 
\begin{cases}
  2 a_{1j} \cos ( k_\ast x - \theta_j),   & x\to -\infty \\
  2 a_{2j} \cos ( k_{\ast} x + \theta_j),   & x \to +\infty.
\end{cases}
\]
Since $w_j$ and $\mbox{Re}(w_j)$ satisfy  the same Helmholtz equation and the same
asymptotic conditions  for $x \to \pm \infty$, we can replace $w_j$ 
by $\mbox{Re}(w_j)$. 

If the structure is symmetric in $x$, then the reflection coefficients
for left and right incident waves are the same, thus 
\[
{\bm S}  = \left[ \begin{matrix} R & T \cr T & R \end{matrix} \right].
\]
Then $\lambda_1=R+T$, $\lambda_2 = R-T$, and 
\[
{\bf a}_1 = \frac{1}{\sqrt{2}} \left[ \begin{matrix}
    1  \cr 1 \end{matrix} \right], \quad 
{\bf a}_2 =  \frac{1}{\sqrt{2}} \left[ \begin{matrix}
    1  \cr -1 \end{matrix} \right].
\]

\end{document}